\documentclass[a4paper,fleqn,usenatbib]{mnras}

\usepackage{newtxtext,newtxmath}
\usepackage[T1]{fontenc}
\usepackage{ae,aecompl}

\usepackage{graphicx}	% Including figure files
\usepackage{amsmath}	% Advanced maths commands
\usepackage{amssymb}	% Extra maths symbols
\usepackage{color}
\newcommand{\gc}{$\gamma$\,Cas}

\newcommand{\xr}{X-ray}

\newcommand{\xmm}{XMM}

\def\gtrsim{\mathrel{\hbox{\rlap{\hbox{\lower4pt\hbox{$\sim$}}}\hbox{$>$}}}}
\def\ltsim{\mathrel{\hbox{\rlap{\hbox{\lower4pt\hbox{$\sim$}}}\hbox{$<$}}}}
\definecolor{gray}{rgb}{0.5,0.5,0.5}

\usepackage{soul}
\usepackage[usenames,dvipsnames]{xcolor}
\usepackage[normalem]{ulem}

\title{Soft and hard X-ray dips in the light curves of $\gamma$ Cassiopeiae}

\author[M. A. Smith \& R. Lopes de Oliveira]{
M. A. Smith$^{1}$\thanks{E-mail: myronmeister@gmail.com }
and R. Lopes de Oliveira$^{2,3,4}$ \\
$^{1}$National Optical Astronomy Observatory, 950 N. Cherry Ave., Tucson, AZ, 
USA\\
$^{2}$Center for Space Science and Technology, University of Maryland, Baltimore County, 1000 Hilltop Circle, Baltimore, MD 21250, USA \\
$^{3}$Departamento de F\'isica, Universidade Federal de Sergipe, Av. Marechal Rondon, S/N, 49000-000 S\~ao Crist\'ov\~ao, SE, Brazil \\
$^{4}$Observat\'orio Nacional, Rua Gal. Jos\'e Cristino 77, 20921-400, Rio de Janeiro, RJ, Brazil \\
}
\date{Accepted WWW. Received YYY; in original form ZZZ}

\pubyear{2019}

\begin{document}
\label{firstpage}
\pagerange{\pageref{firstpage}--\pageref{lastpage}}
\maketitle

\begin{abstract}
The available six archival {\it XMM-Newton} observations of the anomalous 
X-ray emitter \gc~(B0.5\,IVe) have been surveyed 
for the presence of soft X-ray ``dips" in X-ray light curves.
In addition to discovering such events in the soft-band ($\le$ 2\,keV),
we show that sometimes they are accompanied by minor, nearly simultaneous dips 
in the hard X-ray band.  Herein we investigate how these occurrences can be 
understood in the ``magnetic star-disk interaction" hypothesis proposed in
the literature to explain the hard, variable X-ray emission of this Be star.
In this scenario the soft X-ray dips are interpreted as transits
by comparatively dense, soft X-ray-absorbing blobs that move across
the lines of sight to the surface of the Be star.
We find that these blobs have similar properties as the ``cloudlets" 
responsible for {\it migrating subfeatures} in UV and optical spectral lines 
and therefore may be part of a common distribution of co-rotating occulters. 
The frequencies, amplitudes, and longevities of these dips vary widely.
Additionally, the most recent spectra from 2014 July suggest that the ``warm" 
($kT$ $\approx$ 0.6-4\,keV) plasma sources
responsible for some of the soft flux are much more widely spread over the 
Be star's surface than the hot plasma sites that dominate the flux at all 
X-ray energies.  We finally call attention to a sudden drop in all X-ray 
energies of the 2014 light curve of $\gamma$\,Cas and a similar sudden
drop in a light curve of the ``analog" HD\,110432.
We speculate that these could be related to appearances of particularly 
strong soft X-ray dips several hours earlier. 

\end{abstract}

%%%%%%%%%%%%%%%%% BODY OF PAPER %%%%%%%%%%%%%%%%%%

\section{Introduction}
\label{intrdn}

$\gamma$\,Cas (B0.5\,IVe) is an enigmatic hard, variable X-ray star.
Spectroscopic analysis of its emission shows that 
it consists of 3-4 components of optically thin thermal plasma 
\citep[e.g.,][hereafter ``SLM12a''] {Smith et al.2012a}. 
Some 85-90\% of this flux is emitted by a 
``hot" plasma with energy temperature  $k$T $\approx$ 14\,keV, an energy
that cannot be attained by wind-driven shocks or infall onto the Be star. 
The remaining 10\% or so is emitted by mainly  ``warm plasmas" with $kT$ 
of $\approx$0.6-4\,keV.

\gc\ is no longer a unique case of an active Be/X-ray star. It is the 
prototype, and still the brightest member, of a growing class of Be/X-ray stars.
New members have been identified from cross-correlations of optical/infrared 
and X-ray catalogs, typically from sources 
distinguished by their anomalously high X-ray and infrared fluxes and colors 
\cite[see e.g.][]{LO06,Nebot Gomez-Moran et al.2015,Naze & Motch2018}.
These stars have spectral types in the
range of about O9.7--B1.5 and a luminosity class of IV, indicative
of advanced main sequence ages.
\citet[][``SLM16'']{Smith et al.2016} have reviewed the X-ray findings 
through 2015 for this class, now numbering 20$+$ members. 
However, as we will see, recent observations of $\gamma$\,Cas continue to
reveal surprises not found in other types of X-ray emitting, massive stars.  

$\gamma$\,Cas itself is widely taken to be a near-critical rotator, with a
rotational period P$_{\rm rot}$ of 1.21585 days \citep[][``S19'']{Smith et 
al.2006, Smith2019}.
It is well known to be in a wide binary (P$\approx$203.6 days)
with a nearly circular orbit \citep[][SLM12a]{Nemravova et al.2012}.
The evolutionary status of the companion is unknown, but its 
mass is 0.9${\pm 0.4}$ M$_{\odot}$. 
Cross-correlations of {\it International Ultraviolet Explorer 
(IUE)} spectra have established that the contribution 
of the secondary to the \gc~ system's UV flux is $<$0.6\% 
\citep[][]{Wang et al.2017}.
This limit demonstrates that the
faint companion cannot produce the observed rapid UV flux variations found
in the $\gamma$\,Cas system, unless it is an sdOB star
\citep[e.g.,][]{Heber2009}.

Two explanations have been put forward to explain the production of hard
X-ray flux in $\gamma$\,Cas and analogs of the class. The first is through 
accretion of disk or stellar wind onto a degenerate companion, with 
subsequent thermalization of the infall. Although very successful in
explaining the emissions of various classes of X-ray Be binaries, there are
difficulties with the accretion picture explaining them for $\gamma$\,Cas
stars \citep[][SLM16]{Motch et al.2015}.
These include the correlations between  X-ray and UV fluxes in March 1996
and the changes in X-ray color accompanying variations in H$\alpha$
emission in the $\gamma$\,Cas analog, HD45314 \citep{Rauw et al.2018}.
%The correlation of this optical activity with X-ray temperature in 
%particular is hard to reconcile with infall onto a faint companion.  

The second explanation for hard X-ray generation is the ``magnetic 
interaction" hypothesis. As the name implies, in this picture two magnetic 
field systems, one from the star and the other from its decretion disk,
influence one another. The star's magnetic system is thought to arise 
from the development of short-lived magnetic fields generated by subsurface 
convection \citep[][]{Cantiello et al.2009,Cantiello&Braithwaite2011}.
The second magnetic system is a toroidal field created by 
a presumed global Magneto-Rotational Instability in the Be disk 
\citep[][]{Balbus&Hawley1991}.

 \citet[][``H16"]{Hamaguchi et al.2016} have reported time-variable, soft
X-ray ``dips" in X-ray color curves of $\gamma$\,Cas from {\it Suzaku} data.  
They have proposed that X-ray active centers are powered by accretion
onto the surface of a putative white dwarf (WD) companion by a dense Be
wind.  In their picture these centers are occasionally occulted by wind 
clumps as they fall onto the WD, causing absorption of soft X-rays.
In this paper we explore this phenomenon in the context of the 
magnetic interaction picture.

\section{{\it XMM-Newton} observations of \gc}
\label{xrlc}

Launched in 1999 December 10 by the {\it European Space Agency}, XMM 
is an X-ray space observatory that has been in nearly continuous observation.  
Its {\it European Photon Imaging Cameras (EPIC)}, comprised of three
independent detectors (MOS1, MOS2, and pn),
have maximum sensitivity in the energy range 0.3-12\,keV and are used to provide images, 
medium resolution spectra, and relatively high time resolution light curves.
In addition, two other independent instruments onboard XMM are the {\it Reflection Grating Spectrometers (RGS1 and RGS2)}. 
Operating at low energies (0.3-2.5\,keV), the RGSs provide high
resolution spectra and (lower sensitivity) light curves. 
This work was based on archival XMM-{\it Newton} observations of 
$\gamma$\,Cas obtained from its science archive (XSA), which were carried 
out at three different times (Table\,1):

\begin{enumerate}
\item on 2004 February 5, in a 
a single observation with the $pn$, only among the EPIC cameras, 
%(no MOS1 or MOS2) 
running in the {\it fast timing mode}, and with both RGS cameras;
\item in 2010 with observations by all cameras in 
four visits, each separated from the next by about 2 weeks:
July 7, July 24, August 2, and August 24 (investigated and denoted as 
2010-1, 2010-2, 2010-3, and 2010-4, respectively, by SLM12a), and 
\item on 2014 July 24, with all X-ray cameras. 

\end{enumerate}

\begin{table}
\centering
\caption{Log of {\it XMM} observations and the corresponding orbital 
binary phases of the $\gamma$\,Cas system}
\begin{tabular}{@{}lcccc@{}}
  Year &  RJD  &  Duration (ks) &  Obs ID &  $\phi$$_{\rm orb}$ \\
\hline
2004 &    53041.22 &    71   &     02012201  &   0.51 \\
2010-1 & 55384.30  &    18   &     06516702  &   0.74 \\
2010-2 & 55401.15  &    16   &     06516703  &   0.81 \\
2010-3 & 55410.77  &    18   &     06526704  &   0.86 \\
2010-4 & 55428.09  &    24   &     06516705  &   0.95 \\
2014   & 56863.06  &    34   &     07436001  &   0.26 \\
\hline
\end{tabular}
\end{table}
%NOTE:  respective Ph_rot of 2010 are 0.68, .54, .45, .70
%Phases for 2 & 4 are nearly the same; for 1 & 3 overlap of 8ks in phase

All these archival data were reduced anew and processed with SAS V17.0.0 tools by using calibration files available in 2018 December 21. 
The tasks \textsc{epproc} and \textsc{rgsproc} were initially applied to 
the EPIC and RGS data, respectively. The data processing followed the usual 
way applied to XMM-{\it Newton} observations, but assuming a conservative 
approach to avoid pile-up in the EPIC data excluding those collected within 
$<$ 10\,arcsec from the position of $\gamma$\,Cas in the images. 
 
 All spectra were binned such that each energy bin contains at least 25
counts, reducing Gaussian errors to a reasonable level.
We used tools in XSPEC v12.10.1f to analyze the spectra. We compared these
spectra with theoretical models by optimizing the $\chi^2$ criterion as 
fit and test statistic. 

Except for the 2004 observation, for which only the XMM/pn data were 
available among the EPIC cameras, the pn, MOS1, and MOS2 light curves of 
all other observations were merged to compose their final, best quality 
light curves. We constructed the EPIC light curves in different energy bands, 
as described in the text. We followed the procedure of 
\citet[][``LSM10"]{Lopes de Oliveira et al.2010} for the 2004 XMM/pn data.
These were obtained in the fast timing mode and thus have a uncertain 
calibration and increased noise at low energies. Therefore, we did not 
consider the pn data with energies below 0.8\,keV 
for this observation, and have verified the timing analysis at low energies 
by investigating RGS light curves. We merged fluxes in orders 1 and 2 for both
RGS1 and RGS2 to generate a light curve in the energy range of 0.8-2\,keV,
so as to check the behavior of \gc\ in low energies for the 2004 observation. 

Although not relevant for the purpose of this work, times were corrected 
to the barycenter of the solar system. In all cases we used light curves 
with initial time bins of 10\,s.  \textsc{xronos} v5.22  was used for 
exploration of light curves. Both tools, \textsc{xspec} and {\textsc xronos}
are available as NASA/GSFC's HEAsoft software v6.26.

\section{Results of archival {\it XMM} observations}

\subsection{Light and soft/hard color curves for three {\it XMM} epochs} 
\label{xdlc}

 The {\it XMM} data were examined over various 
combinations of energy bands in the energy range 0.3-10\,keV. The presence 
of certain ``soft dip" (SD) features in the color ratio curves 
were noticed in various combinations of soft and hard energy bands; these
will be discussed below.  Individual soft bands are specified in the figures.
The ``hard" X-ray band is taken as the 2-10\,keV, unless otherwise noted.

\subsubsection{The 2014 light curve}
\label{2014lc}

Figure\,1 exhibits the hard X-ray curve (2-10\,keV; solid line) and a 
soft/hard color ratio curve (0.8-2\,keV/2-10\,keV) 
for the 2014 time series. 
The figure shows three probable semi-resolved soft/hard X-ray dips 
(``SD") in this interval. Because they are only semi-resolved, we will
generally refer to them as a single aggregate feature in this paper. The
central region of the aggregate may or may not be flat.
This dip is flanked by a steep decrease of only about 25\% of the original 
soft/hard energy color, followed by a similar steep rise to the initial color. 
The weak response in a similar {\em hard} band feature suggests that 
these color curve events are mainly due to soft flux variations.

An interesting aspect of the SD complex is a slight delay relative to the 
centroid of the associated dip in {\em hard flux} at 1.5-2.5 ks. We found this 
delay, visible as the small time difference between the two vertical lines 
to the left in the figure, to be 175${\pm 60}$\,s. 
This value was obtained by measuring 
the centroids of least-squares Gaussians of the two features by means of 
the {\it Interactive Data Language} routine, {\em gaussfit.pro}. It merits
report primarily  because of similar SD delays found in 2004 data 
(see $\S$\ref{2004lc}). 

%FIGURE1
\begin{figure} 
%\label{gc2014lc}
\begin{center}
\vspace*{0.01in}
\includegraphics*[width=8.45cm,angle=90]{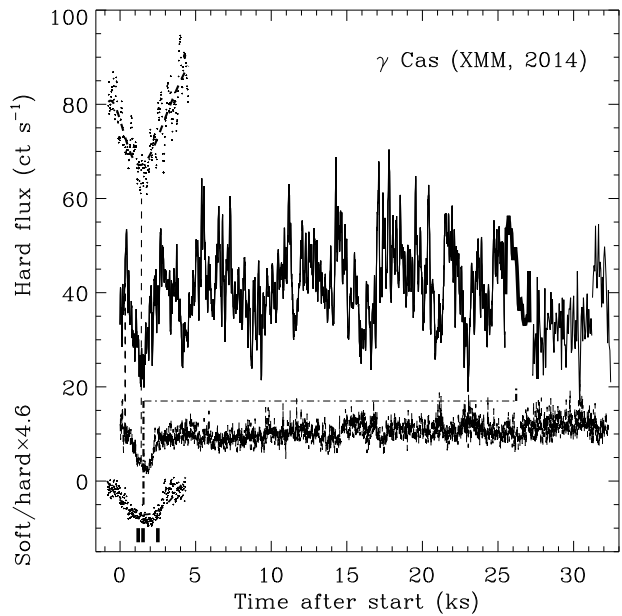}
\end{center}
\caption{ 
Upper and lower panels exhibit the hard \xr\,\,and soft/hard 
color curves for \gc\,\,during the 2014 July 24 {\it XMM} observation. 
Times are rebinned to 50\,s.  A short vertical lines at 1.5-2.5\,ks 
above the X-ray color curve connected to a dotted
dashed line depict three marked semi-blended soft (0.8-2\,keV) X-ray ``dips." 
The dips in the hard flux and color ratio curves are enlarged in time
by a factor of 2; Gaussian fits are superimposed.
The right end of the horizontal dashed light connect to a ``hard flux drop" 
during times 26.7-31.2\,ks.  The start time is modified from that of Smith 
et al.  (2012a) to designate simultaneous camera operations.
}
\end{figure}

  A second major feature in Fig\,1 is the sudden drop by $\approx$35\% in 
the X-ray fluxes after time 25\,ks, except for a likely ``quasi-flare" 
aggregate at 32\,ks. 
We cannot state with certainty that the drop's presence at the end of the 
time series is only the beginning of a long-lasting, new flux plateau.  
However, it is a unique feature among the published X-ray light curves 
of \gc.~ For example, it is more than five times longer than brief, 
broad-band X-ray ``cessations" discussed by 
\citet[][``RSH'']{Robinson & Smith2000},
\citet[][``RSH02'']{Robinson et al.2002}, and LSM10. 
We notice also in Fig.\,1 that the drop in the soft energy fluxes 
appears to be slightly less than the hard fluxes, a point verified in 
our spectral analysis ($\S$\ref{modspec14}).

\subsubsection{The 2010 light curves}
\label{2010lc}

%FIGURE2
\begin{figure} 
%\label{gc2010lc}
\begin{center}
\vspace*{0.00in}
\includegraphics*[width=8.45cm,angle=90]{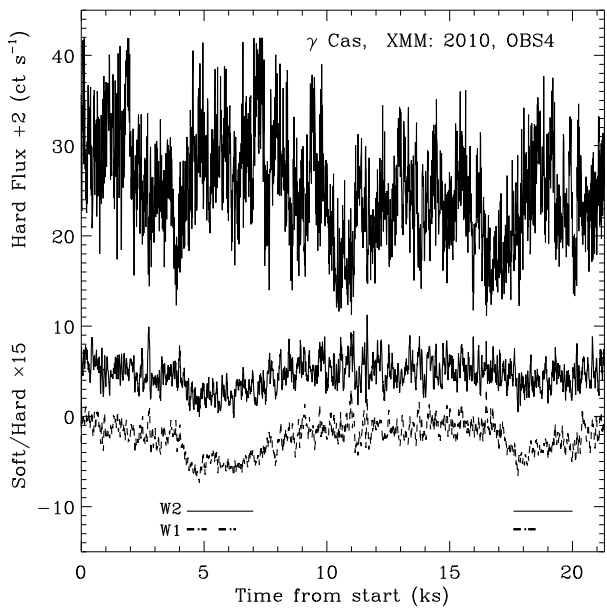}
\end{center}
\caption{
The 2010-4 (August 19) 
light and soft/hard curves (rebinned to 50\,s), where ``soft" is shown
over 0.8-2\,keV (dashed; also called ``moderately soft" in the text) 
and ``very soft" over 0.3-1\,keV (solid). 
Note two soft dip features with minima at 5-7\,ks and 18-20\,ks. 
Time windows W1 and W2 are depicted for separate spectral analyses. For
clarity offsets in y of -10 and -8 units are given for W1 and W2 curves}.
\end{figure}

Figure\,2 exhibits a hard X-ray flux curve for 2010-4. 
In this case we include two soft/hard color curves, with one formed from 
``moderately soft" (0.8-2\,keV) and a second from ``very soft" (0.3-1\,keV) 
fluxes; no corresponding hard flux drops are present. 
The main features of these color curves are preserved when the fluxes 
are taken from the RGS for comparison purposes rather than the EPIC 
cameras actually shown. According to the moderately soft color curve, 
time slices of the dips at 4.3-5.1\,ks, 5.6-6.6\,ks, and
17.6-18.3\,ks exhibit almost flat-bottomed profiles. This is to say,
the soft dips consist of unresolved and nearly resolved features.
These windows will be considered separately for broader time intervals at
4.3-7\,ks or 17.6-20\,ks in our spectroscopic treatment ($\S$\ref{spmodl}).

 SLM12a found no SDs in 2010-1 or 2010-3. However, in the 2010-2 observation 
they found six brief, shallow SDs (three separated by 2\,ks, and three 
more by 1\,ks). Again these were evident in the soft fluxes either from EPIC 
or, again for checking purposes, RGS cameras. Since there is less
information in them, we will not attempt to quantify the implied column 
densities.  

Adopting S19's ephemeris for the Be star, the rotational phases 
% (referenced to $\phi$ = 0 at RV maximum) 
at the start of the four observations in 2010 are 0.68, 0.54, 0.45, and 0.70. 
Note that the phases of 2010-1 and 2010-4 are essentially the same, 
i.e., there is substantial overlap of the implied visible stellar longitudes.
Thus, whereas there are two SDs present in 2010-4, there are none in
2010-1. This means that the structures responsible for the soft X-ray
occultations develop in less than the time interval between them,
44\,days, and possibly much less. Additionally, the phase overlap 
between 2010-2 and 2010-3 is still sufficient to determine that none of
the six SD events in 2010-2 were present in 2010-3, 9 days later.
These are the first instances in which we can establish upper limits
to the longevity of specific SD events.

\subsubsection{The 2004 light curve }
\label{2004lc}

A search for soft/hard color dips within ${\pm}$1\,ks of the hard flux dips 
in the 2004 light curve disclosed five short-lived ($\ltsim$1\,ks) soft/hard 
dips, which as with the previous datasets are strongly present in the soft
X-ray curves alone. The significance floor for these color events was at least
%The color ratio dips were identified and found to be significant to at least 
3$\sigma$,\footnote{The significance levels were computed by a Monte Carlo-type
simulations in a  program, {\it ewcalc.pro,} described by \citet{Smith2006}. 
This program uses the fluctuation level derived 
in the flat sections of a time series or spectrum to compute the rms noise 
level. It then calculates the number of random realizations required to 
equal the ``equivalent width" of an apparent feature.}
As for the events in 2014, for 2010, all five of these features were present 
for various combinations of hard and soft band energies, including RGS data.
The soft-dips, though sharp, have soft/hard amplitudes of 20-25\% 
relative to the hard fluxes. These dips appear to be {\em nearly}
simultaneous with small hard flux dips but,
like the longer-lived dip complex in the 2014 data, exhibit small lags of 
180-600\,s.  The upper panel of Figure\,3 shows the full hard flux and 
soft/hard color time series, while the lower panel displays enlargements
of the five SD events.

%FIGURE3
\begin{figure} 
%\label{gc2004lc}
\begin{center}
\vspace*{-0.00in}
\includegraphics*[width=8.45cm,angle=90]{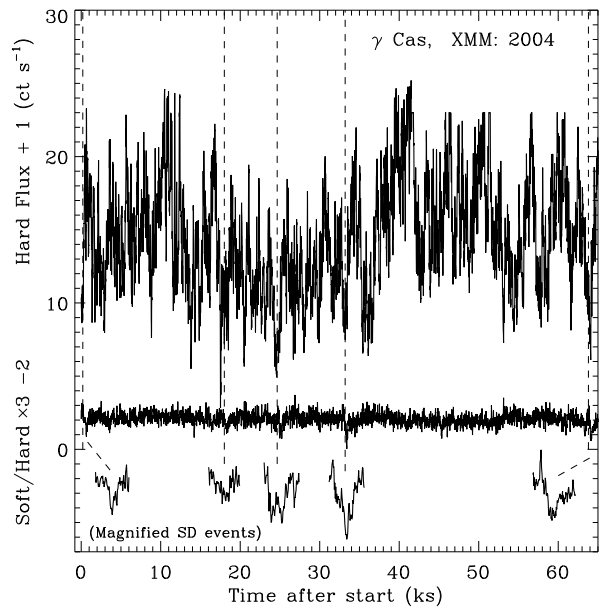}
\end{center}
\caption{ 
Hard flux (upper0 and soft/hard color (lower) 
curves for the 2004 February 5 observation,
where ``soft" is taken again as 0.8-2\,keV.  Associated with five hard flux 
dips (vertical dashed lines) are brief color (SD) dips delayed by 200-500\,s. 
Individual events are magnified below (2$\times$ in color, 5.5$\times$ 
in time.) Offsets in y are given for the sake of visualization }
\end{figure}

  To summarize, our analysis of low resolution spectra demonstrates that 
SDs found in 2014, 2010-4, and 2004 are similar in general morphology 
but differ in details such as their amplitudes, sharpness, and longevity.

\subsection{{\it XMM}/EPIC Spectra }
\label{xrsp}

The soft and hard X-ray light curves of $\gamma$\,Cas 
generally show similar variations (e.g., LSM10). 
However, we found occasions 
%especially in the 2014 and 2010 (OBS2 and OBS4) datasets, 
during which flux curves for ``soft" and ``hard" behaviors were quite different.
Also, we noted an extended ``hard flux drop" (HD) in the 2014 observation.  
The times for the soft dips and ``hard drop" in Fig.\,1 are 0.8-2.2\,ks
and 26.7-31.2\,ks, respectively.  We proceeded to optimize their different 
behaviors by experimenting with soft and hard energy spectra 
with different soft-energy ranges. 
This step enabled us to define a convenient demarcation 
energy of 2\,keV and thus the abutted limits of their energy ranges.
In like manner we contrasted soft dip fluxes from the 2010-4 curve 
with fluxes from non-dip intervals.

We emphasize that it is beyond the scope of this work to proceed with 
a detailed spectral analysis. In the following, we will 
evaluate the XMM/EPIC spectra to explore the hypothesis that the variabilities
reported in this paper are due to changes of the primary X-rays generated near
the Be star and modified by changes in the local absorber toward the observer.
To do this, we attempted to apply the model that SLM2012a determined from the 
2010 observations again to the 2010-4 spectrum and also to spectra of time
slices of the 2014 light curve. However, we found that the
conditions during selected times of the 2014 observation varied too much
to permit as good a fit as found for the 2010 data by SLM12a. This was
particularly true for the description of the low energy tail, then resulting
in dubious values for the emission measures of the coolest components kT$_1$
and kT$_2$.

\subsubsection{Modeling the spectrum of 2010-4 }
\label{spmodl}

Here we revisit the 2010 XMM-{\it Newton}/EPIC spectra already investigated 
by SLM12a under a new approach. The aim is to disentangle spectral features 
that characterize the soft and hard photometric dips reported in this work, 
which are marked in Fig.\,4 as 
W1 (4.3-5.1, 5.6-6.6, 17.6-18.3\,ks) and W2 (4.3-7, 17.6-20\,ks), respectively. 
In this exercise we extracted three sets of spectra, each one including MOS1, 
MOS2, and pn spectra: the first set corresponds to data collected during W1, 
the second during W2, and the third, namely ``nominal'', from the remaining
times (precisely, $<$4\,ks, 8.5-17\,ks, and $>$ 20.2\,ks).
%between the two -- TT-UU\,ks in Fig. 4. 
For the description of these spectra, 
we use the model presented by SLM12a that resulted from a comprehensive 
spectral analysis that included high resolution spectroscopy from RGS cameras.
%Because of the relatively short duration of W1 and W2 we did not
%add RGS fluxes to the 2014-4 analysis, except as a qualitative check on 
%the reality of these dips. 

The model then explored is that of SLM12a: 
$n_{H_a}*(T_1+T_2+T_3+T_4)+n_{H_b}*(T_4+gauss)$.  
Our fitting procedure was to create one group in XSPEC for each sets of 
spectra, respectively.
These were used in a simultaneous fit with a common statistics for the nine 
spectra. The distinction in fits for an individual group was that each of 
them had $n_{H_a}$ and normalization factors of the thermal components 
varying independently of the values of the other two groups. However, the 
temperatures ($k$T\,=\,0.11\,keV, 0.62\,keV, 3.43\,keV, and 13.51\,keV) and 
$n_{H_b}$ values (73.7$\times$10$^{22}$ cm$^{-2}$) were kept fixed to the 
values already obtained by SLM12a. 
The fit resulted in$\chi^{2}_{\nu}$\,=\,1.10
(degrees of freedom of 4602). In units of 10$^{22}$ cm$^{-2}$, and with 
uncertainties of ${\pm 0.01}$ at 1$\sigma$, the results for 
n$_{H_a}$ were 0.10, 0.23, and 0.20 for the ``nominal'', W1, and W2 interval 
spectra, respectively. The emission measures\footnote{We corrected a 
typographical error in SLM12a's Table\,4 that made the EM for plasma \#2 
appear 10$\times$ too large. We thank H16 for pointing this out.} and 
absorption values are presented in Table\,2.

\begin{table*}
\begin{center}
 \caption{2010-4: best-fit spectral parameters from model $(T_1+T_2+T_3+T_4)*n_{H_a}+(T_4+gauss)*n_{H_b}$ of Smith et al. 2012a.}
\label{tbl:obs4_2010}
\begin{tabular}{llllllll}

\hline
\hline
%\tableline
%\tableline
State   & n$_{H_a}$ & n$_{H_b}$  & EM T$_{1}$ ($\times$10$^{55}$\,cm$^{-3}$) &  EM T$_{2}$        & EM T$_{3}$        & EM T$_{4,Nha}$    & EM T$_{4,Nhb}$    \\
          & (10$^{22}$\,cm$^{-2}$) & (10$^{22}$\,cm$^{-2}$)  & ($k$T\,=\,0.11 keV)     & ($k$T\,=\,0.62)           & ($k$T\,=\,3.43)          & ($k$T\,=\,13.51)         & ($k$T\,=\,13.51) \\    
%\tableline 
%\tableline
%\hline
\hline
Nominal window & 0.10$^{+0.01}_{-0.01}$ & 73.7$^1$ & 0.14$^{+0.04}_{-0.03}$ & 0.04$^{+0.01}_{-0.01}$ & 0.32$^{+0.02}_{-0.02}$ & 2.65$^{+0.02}_{-0.02}$ & 0.90$^{+0.14}_{-0.14}$ \\ 
W1 window       & 0.23$^{+0.01}_{-0.01}$ & 73.7$^1$ & 0.13$^{+0.14}_{-0.12}$ & 0.04$^{+0.01}_{-0.01}$ & 0.32$^{+0.07}_{-0.07}$ & 2.92$^{+0.06}_{-0.06}$ & 0.69$^{+0.34}_{-0.34}$ \\
W2 window      & 0.20$^{+0.01}_{-0.01}$ & 73.7$^1$ & 0.24$^{+0.10}_{-0.09}$ & 0.04$^{+0.01}_{-0.01}$ & 0.36$^{+0.05}_{-0.05}$ & 2.86$^{+0.04}_{-0.04}$ & 1.03$^{+0.24}_{-0.24}$ \\ 
\hline

\end{tabular}
\end{center}
%\item 
{\bf Notes:} Temperatures and n$_{H_b}$ from Smith et al. 2012a. Abundances solar, except $Z_{FeK\alpha}$ = 0.18$\times$$Z_{Fe, \odot}$, $Z_N$ = 2.33$\times$$Z_{N, \odot}$, $Z_{Ne}$ = 1.8$\times$$Z_{Ne, \odot}$. \\

\end{table*}

%%FIGURE4
\begin{figure} 
%\label{2010spec}
\begin{center}
%\vspace*{0.20in}
\includegraphics*[width=5.5cm,angle=-90]{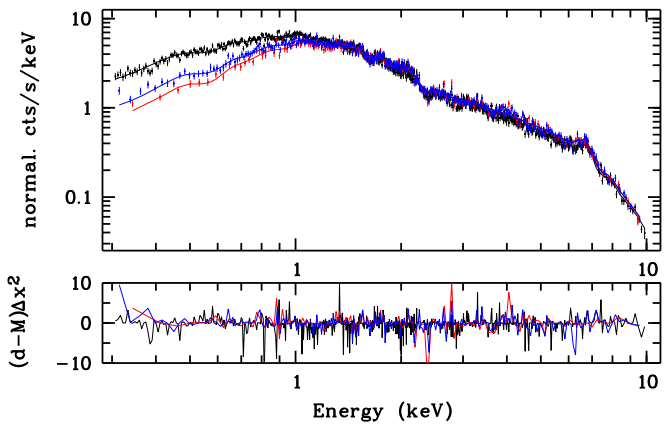}
\end{center}
%\vspace*{-0.15in}
\caption{The 2010-4 {\it XMM}/pn spectra of \gc\,\,plotted according to the 
soft/hard dips and W1, W2 windows shown in Figure\,2.  
Nominal fluxes (displayed as black) are extracted 
from time windows $\le$4.0\,ks, 8.5-17\,ks, and $\ge$20.2\,ks.
W1 (in red) and W2 (in blue) refer to spectra extracted from the 
restrictive, and broader SD time intervals, respectively. 
The lower panel shows the difference between the data and the model fits 
(solid lines).
}
\end{figure}

  We note in passing for Fig.\,2 that the absence
of clear markers of individual SD subfeatures within the W windows,
which would help define saturation levels and lifetimes of unresolved SD
aggregates, suggests that our spectroscopic analysis for 2010-4 is
bound to be somewhat oversimplified: our computed parameters represent
some ill-defined average of possibly several events. Nonetheless, we believe
our n$_{H_a}$ values in Table\,2 are not much different from those of
any unresolved rapid events within the W windows. 
%Also, as established in $\S$\ref{modspec14}
%they are about an order of magnitude lower 
%than the value found for the SD in the 2014 observation. 
Coincidence or not,
notice that the relatively low column densities derived for these events 
are roughly offset by their comparatively long durations.

\subsubsection{Modeling the 2014 spectrum} 
\label{modspec14}

As presented in $\S$\ref{2014lc}, the variability in the brightness of 
$\gamma$\,Cas was remarkably energy-dependent in two events during 2014: a
strong soft/hard flux dip early in the time series,
followed by a relatively strong dip especially in the hard energy fluxes near
the end. Although a detailed spectral analysis is not warranted, we have 
replicated the same reasoning applied to the 2010-4 spectra in $\S$\ref{spmodl} 
to test the hypothesis that the color variations 
during the soft dip are due to changes in the
local X-ray absorbers. Also, we expand the exercise to search for clues 
concerning the origin of the {\em hard drop.} We start by displaying XMM/EPIC 
spectra associated to the soft (SD; 0.8-2.2\,ks) 
and hard (HD; 26.7-31.2\,ks) dips, and a third set corresponding to 
a ``nominal'' state (all other tmes before 25\,ks). For the sake 
of clarity, we show in Fig.\,5 only the XMM/pn spectra of each window.

Departing from the approach applied to the 2010-4 set of spectra, 
we investigate separately the three set of EPIC windowed spectra of the 
2014 observation. First, we applied the same model as for the 2010-4 
observation ($\S$\ref{spmodl}) for the 2014 set of spectra representing 
the ``nominal'' state. We fixed the temperature as for the 2010-4 and allowed 
the $n_{H_a}$, $n_{H_b}$, and normalization of the thermal components to vary. 
It resulted in $n_{H_b}$\,=\,23.7$^{+2.2}_{-1.8}$$\times$10$^{22}$\,cm$^{-2}$ 
and $\chi^{2}_{\nu}$\,=\,1.25 (d.o.f.\,=\,2649). The $n_{H_b}$ value was kept
fixed when applying the same model to the SD spectra 
($\chi^{2}_{\nu}$\,=\,1.21; d.o.f.\,=\,600), then to the HD spectra 
($\chi^{2}_{\nu}$\,=\,1.06; d.o.f.\,=\,1504). The resulting $n_{H_a}$ values 
in units of 10$^{22}$\,cm$^{-2}$ were 0.18$\pm$0.01, 0.87$\pm$0.02, and 
0.12$\pm$0.01, for the ``nominal,'' SD, and HD sets of spectra, respectively.
Thus, the $n_{H_a}$ value for the SD in 2014 is some four times larger than 
the maximum value we found for 2010-4. (However, note that the SD spectrum 
refers to the entire event, including ingress and egress times, so this column 
may be a lower limit.) Interestingly, the $n_{H_a}$ for the
the HD is actually less than for the nominal fluxes. This is consistent
with the very slight rise in the color in Fig.\,1 compared to the level 
earlier in the time series.

%%FIGURE5
\begin{figure} 
%\label{2014spec}
\begin{center}
%\vspace*{0.20in}
\includegraphics*[width=5.5cm,angle=-90]{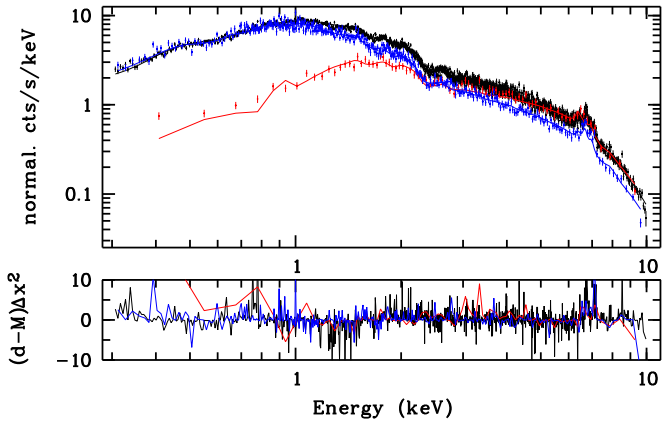}
\end{center}
%\vspace*{-0.15in}
\caption{The 2014 {\it XMM}/pn spectra of \gc\,\,plotted according to soft \xr\
dip (``SD,"in red), hard \xr\ {\em drop} (blue), and ``nominal flux" (black)
times from Figure\,1. 
}
\end{figure}

 Notice in Fig.\,4 and Fig.\,5 that the ```ultra-soft" 
energies ($\le$0.8\,keV) extracted from the 2014 nominal flux times
and the fluxes from the W1 and W2 times (2010) no longer diverge because
the spectral slopes become parallel.
In fact for the 2014 data (Fig.\,5) the observed fluxes at
$<$0.8\,keV exceed the predicted ones, even though the match is quite
good for less soft fluxes at 0.8-2\,keV. We discuss this behavior, 
also found in H16's {\it Suzaku} spectra, in $\S$\ref{sftbehav}.

 We digress to point out that although the major thrust of this paper is to
study rapid changes in X-ray absorption in \gc,~ i.e., $n_{H_a}$, our derived
value of n$_{H_b}$\,=\,23.7$\times$10$^{22}$\,cm$^{-2}$ is interesting
in the larger context of this star's outburst history. SLM12's {\it XMM} 
2010 observations and 2010-2011 optical continuation showed that the star
had just started a Be outburst, resulting in matter being added to the disk
and perhaps wind. At that time n$_{H_b}$ increased over the range
36-74$\times$10$^{22}$\,cm$^{-2}$. Our 2014 value shows that this
activity was still ongoing. By comparison, in a 2004 observation LSM10 
found a $n_{H_b}$ of only 0.23$\times$10$^{22}$\,cm$^{-2}$.

\section{Discussion}

\subsection{The context of the magnetic interaction picture}
\label{cntxt}

As noted in $\S$1,
\citet[][]{Cantiello et al.2009} and \citet[][] {Cantiello&Braithwaite2011} 
have predicted that
%the magnetic interaction idea, 
an iron-opacity convection zone is set 
up within the outer equatorial zone of very rapidly rotating massive stars.  
These authors have suggested
that the convection should generate small-scale magnetic fields,
the strengths and scale lengths of which are characterized by the convective 
velocities and cell sizes. 

Moreover, the tangling of individual field lines from the stellar surface and
toroidal lines from the Be disk causes stresses and ultimately their breaking
and reconnections.  The ensuing relaxation of the magnetic tensions proceeds 
as a slingshot, thereby accelerating high energy electron beams to the 
surface of the Be star. 
%For a more complete description see
This description has been elaborated upon by 
\citet[][``RS00'']{Robinson & Smith2000}, 
\citet[][``RS02'']{Robinson et al.2002}, and S19.

The X-ray light curves of $\gamma$\,Cas exhibit ubiquitous rapid 
``quasi-flares" that  \citet[][``SRC98'']{Smith et al.1998a}
interpreted as explosive releases of beam energy in the photosphere. 
Such events of course are not true flares in the solar sense but rather the 
manifestations of
{\em thermal} conversion of the beam energy and the subsequent adiabatic 
expansion of hot plasma, which then accumulates in a magnetically confined,
overhead ``canopy" of lower density. It is in fact in these canopies
where most ($\sim$70\%) of the hard (``basal") X-rays are released. 
This energy is also released as X-rays over a longer timescale as the
canopy plasma cools. Empirically, the flare and basal flux temperatures are 
the same to within $\approx$10\% (SRC98), a fact that sets important 
constraints for physical models.

 S19 have identified the X-ray emitting canopies with 
co-rotating ``cloudlets," observed as the Doppler motion of 
{\em migrating subfeatures (msf)} in UV and optical line profiles. 
These bodies absorb flux as they move transversely across the line of sight
in front of surface activity centers \citep[SRC98,][]{Smith&Robinson1999}.
The {\it msf} have common timescales of appearance and inferred column 
densities as the basal X-ray canopies, and thus they are likely the same 
bodies. or at least parts of a common distribution.
We reiterate that the evidence for cloudlets comes from UV and optical 
line variability. Therefore they must originate from the Be star.
% there is no place for them in the accretion picture, 
% because again faint companions cannot cause this variability.

\subsection{Interpretation of the 2014 hard X-ray dip accompanying 
the soft dip}

To discuss the X-ray dips, and beginning with the 2014 
hard dip,
%\footnote{By ``hard dips," we distinguish between minor decreases
%in the hard X-ray light curves from the ``hard {\em drops}" discussed
%depicted in Figures\,1 and 6.}
we consider, first, that the magnetic strengths associated with the 
convective cells predicted by Cantiello et al. have a limit imposed by
the equipartition of their magnetic and turbulent energies.
Accordingly, an ensemble of active cells across the surface of the 
Be star will host a distribution of magnetic strengths peaking at 
this limit and tapering to lower values. Although most cells should
have this characteristic magnetic field strength, an 
occasional one will host a weaker one.

As a second consideration, 
we return to the SRC98 picture of the generation of quasi-flares and
the resulting basal flux emitted from an overhead canopy.  Since the source 
of these fluxes is the incident electron beam, the {\em emission measure} of
the flare, EM$_{\rm flare}$, is proportional to the beam energy. However, 
this strict proportionality does not necessarily extend to EM$_{\rm canopy}$ 
as well because of nonadiabatic losses there: first as the canopy fills,
and more importantly as the post-flare canopy plasma cools.
The basal emission is related to the canopy's
magnetic confinement, which in turn depends on both
the magnetic flux at the surface anchor footpoints and 
the gas-magnetic pressure equilibrium at the apex of
the canopy.  Indeed, the abrupt ingress/egress of the soft and hard X-ray 
dips associated with the canopy has sharp leading and trailing edges 
suggests that it is confined by an external pressure.

Turning to a possible explanation of the hard flux dip in Fig.\,1, 
since the {\em emission measure} of the canopy, EM$_{\rm canopy}$ 
$\propto$ $N_{e}^{2}$\,$V_{\rm canopy}$, and given that the volume is 
inversely proportional to the electron density, it follows that
EM$_{\rm canopy}$ is simply proportional to $N_e$.
Then, the overall hard X-ray flux contribution from all 
canopies above active centers on the visible disk depends mainly on
the {\em average} magnetic field surrounding the canopies. 

Elaborating further on this description, let us consider that
whenever an electron beam impacts an surface active center 
which hosts a weaker than average surface field, the radius 
of the associated canopy, such as we will
associate with any of the three SD events in Fig.\,1,
will become larger than others associated with stronger fields. 
Given its relatively low density, the EM of our {\em designated}
canopy will be likewise low, the deficit having gone into more
adiabatic expansion to fill a larger volume. 
The sum of all basal flux contributions 
across the visible disk will now be decreased by a small amount
because of our one canopy's reduced emission. Meanwhile,
several other emitting canopies, each with an average EM, 
will produce greater X-ray fluxes than our designated canopy.
Altogether, its weaker field will result in a reduction 
in the basal flux integrated across the disk.

\subsection{The soft X-ray dips }

\subsubsection{The SDs in the \xmm~ light curves}
\label{sall}

  In the 2014 time series each of the three semi-resolved SDs last
0.5-1\,ks. We interpret their presence as being caused by 
absorption by one or more canopies, 
moving transversely in the line of sight over a small magnetic 
active surface region on the Be star. Since during even the core 
phase, the soft/hard ratio dips only by $\sim$25\%,
the area of the X-ray source on the star is likely smaller 
than the overhead occulting cloudlet. Thus, 
the canopy flanges outward from the active surface area projected toward 
the observer. H16 reached similar conclusions.

In $\S$\ref{cntxt} we identified the primarily UV-absorbing ``cloudlets" with
structures responsible for {\em msfs} appearing across line profiles.  
Combining the inferred transverse velocities of 95\,km\,s$^{-1}$ of these
structures as they pass through the stellar meridian {\em and} their 
lifetimes of $\sim$5\,ks for the
largest feature occurring at 1.5\,ks in Fig.\,1, we find the diameter 
of a typical cloudlet is $\sim$5$\times$10$^{5}$\,km. 
This UV-based timescale is comparable to the 
X-ray based timescale for the
strongest feature(s) occurring at 2\,ks in Fig.\,1. Thus, the derived diameter 
of a typical cloudlet becomes 3-5$\times$10$^{5}$\,km. 
While still small, this value is much larger than the initial size of 
individual quasi-flare parcels, $\sim$3-5$\times$10$^{3}$\,km 
(SRC98), on the star's 
surface.  RS00 showed that a high energy electron beam of this diameter 
directed toward the star can generate a considerable amount of X-ray energy 
(1.5$\times$10$^{32}$ erg) upon impacting the stellar surface. This is of
order 15\% of the integrated energy of a flare on $\gamma$\,Cas.

SRC98 gave an rough particle density estimate of 10$^{11}$\,cm$^{-3}$ for 
a canopy, which we may now compare with an estimate from time delays of 
the SDs seen in the 2004 and 2014 light and color curves. 
The straightforward interpretation of 
the SD lags is that they represent a decay timescale of a plasma with a lower 
density than the initial quasi-flare volume. A decay of $\sim$300\,s is 
expected for a hot plasma with a density of 3$\times$10$^{11}$\,cm$^{-3}$, 
or just a factor of three off the earlier estimate. 
Although neither of these values is precise, they are determined 
by independent means.

\subsubsection{Size comparisons for corotating occulting structures}

  A check on our tentative inference that so-called cloudlets,
responsible for {\it msf} in optical/UV lines, are also responsible for 
at least some of the X-ray SDs 
can be made from estimates of their physical sizes. S19 estimated 
size of typical cloudlets as $\approx$1$\times$10$^{5}$\,km.
If we adopt a time interval of 600\,s separating semi-resolved SD components 
in Fig.\,1 in terms of stellar longitude ranges of rotationally advected blobs, 
next use the star's rotational
period (1.21\,d), radius of 10R${\rm *}$, and finally take a
rotational inclination of 45$^{\circ}$, we can
derive a size of 2-3$\times$10$^{5}$\,km, depending on
whether the active features are at a stellar latitude of 45$^{\circ}$ or 
on the equator.  Perhaps even smaller values obtain
for even briefer SDs, such as suggested by the SDs in 2004 (Fig.\,3). 
However, in such cases, fainter {\it msf} signatures in UV spectra 
might not be observable. 

   If for the X-ray canopy we take these size and electron density estimates, 
and assume a roughly spherical shape, a column density of about 
4.5$\times$10$^{21}$\,cm$^{-2}$ results. This is within a factor of two 
or so of the spectroscopically derived value in $\S$\ref{modspec14}.
Altogether, the physical estimates of the absorbing parameters from 
the X-ray and UV domains are in good agreement.

\subsubsection{SDs in the context of recent literature }

 H16 have reported a few soft (also defined as below 2\,keV) dips 
in a 111\,ks long {\it Suzaku} observation on 2011 July 13-14.
They interpreted these features as arising from absorbing structures 
having column densities of 2.4--8.1$\times$10$^{21}$\,cm$^{-2}$. 
This is similar to the $\sim$10$^{22}$\,cm$^{-2}$ we found for 
the column density responsible for the principal soft dip in Fig.\,1.  
No simultaneous variations in the hard X-ray light curve could be found. 
In all, H16 identified six SD features, probably with time-symmetric 
profiles. In three cases their profiles were ``flat-bottomed," 
indicating saturated absorption, in terms of optical thickness.

The duration and profile of some of the saturated dips, as well as 
the column density H16 derived for an occulting feature moving in front 
of the \xr~ source, reveals the same phenomenology as our 2014 SD event.
The primary difference, crucially, between their interpretation and ours is
that they consider the occulting structures to be Be wind blobs 
moving transversely across X-ray sites on a putative magnetic WD. 

  H16 rejected the idea \xr~ plasma sources could be emitted near 
the Be star for the stated reason that in the interaction scenario active 
centers must be ``spread evenly" across the Be star, whereas these authors 
reasoned that most of the X-ray emission must arise on two centers on 
the \xr-active star. However, other than as a occasional artifice, our
previous papers advocating the magnetic interaction picture were not predicated
on the assumption of a quasi-random spatial distribution of sources. 
(For example, the appearance of large variations of the  basal 
emissions in 1996 January and March are likely due to two widely 
separated major active centers advecting across a rotating Be star 
(SRC98), at least at that time).  Also, 
H16 attributed the major flat-bottomed dips in the {\it Suzaku} data 
to two major X-ray emitting  centers on a magnetic dipole, 
This suggests a full rotational period of roughly 1.6\,days, 
which is consistent with the 1.2\,day period for $\gamma$\,Cas. 
Thus, there is no clear contraindication of the X-ray variations
being emitting from the Be star. Rather, their Dip \#6 is likely a
reoccurrence of Dip \#1 one rotational cycle later if the emissions
originate from the Be star.

 Recently,  \citet[][``H18'']{Hamaguchi2018} has observed SD events 
in $\gamma$\,Cas using the {\it Neutron Star Interior 
Composition Explorer (NICER)} satellite. These observations occurred during
during 2017 June 24 and September 29, for a total duration of 57\,ks, 
This instrument is very sensitive to soft ($\sim$1\,keV) X-rays,
allowing the search for more short-lived color variations.
These light curves show a ``rapid, consecutive [soft X-ray] dipping"
that ``comes and goes" even within 2\,ks. 
These dips are similar to those we reported for the 2004 and 2010 observations.

The occurrence of the SDs distributed in phases around the binary orbit 
(Table\,1) demonstrates that they need not occur during special orbital 
alignments of the primary and secondary with respect to the observer.
Thus there is no evidence for a velocity vector other than transverse motion
for absorbing blobs that cause the SDs.

It is now clear that the foreground occulting canopies/blobs may be 
{\em either} ephemeral and rendered visible sporadically over a particular 
activity center, {\em or} long-lived. In the latter case, lines of sight to
one of them will successively intersect different surface centers that are
advected across the stellar surface by rotation.
Not enough events have really been recorded to establish which is the 
more likely case.  
However, from the limited number of events to date, it can be argued that 
some of them favor one or the other explanations at different times. 
In particular, very closely spaced events in Fig.\,1  suggest
successive activation of small regions over a closely spaced area of the star. 
Alternatively, the series of SDs over a time corresponding to 60\% of the 
Be star's rotational period (Fig.\,3) implies active sites are
distributed over a broad range of surface longitudes.  
Repeated hard/soft X-ray observations showing similar column densities for 
SDs spaced closely in time should further test these possibilities. 

The SD events of observations of \gc~ recorded by H16, H18, 
and those observed by the {\it XMM} ($\S$\ref{xdlc}), 
suggest a picture in which major X-ray centers are not stable. 
This fact contradicts a picture that includes a permanent magnetic dipole, 
regardless of the identity of the X-ray active star. 
It is therefore inconsistent with the presence of a fossil field, 
such as would be required for most magnetic WD accretion models. Our
inference of magnetic impermanence
follows also from an examination of the earlier X-ray monitorings of 
$\gamma$\,Cas. For example, although a dipole picture for the Be star could 
be maintained in principle from the 1996 March observations of SRC98, later 
observations do not bear out a continuity.  Observations spaced a few days
apart in 1998 showed only a trace of similar major X-ray features (RS00). 

Any picture in which hard
X-rays are emitted from a single magnetically active region 
(by necessity, small-scale and highly multipolar) must explain occasional
series of apparent reoccurrences, or ``flickers," of SD events. This 
conclusion is reinforced by the ubiquitous appearance of quasi-flares.
In tallying up the shortcomings of the WD and interaction scenarios, on one 
hand H16's WD scenario suffers the flaws of not accounting for stability 
of robust dipolar fields, and of assuming a very high mass loss rate for
the Be star (see $\S$\ref{mdtt}). Even so, the interaction scenario 
relies on a complicated and unproven interplay between the 
presence of subsurface convective cells and (still undetected) small-scale, 
multipolar magnetic fields that guide high energy electron beams to the surface 
sporadically.

\subsection{Behavior of the soft X-ray fluxes}
\label{sftbehav}

In $\S$\ref{xrsp} we discussed the 2014 and 2010-4) soft X-ray spectra 
and noted two unexpected results. 

The first, in the spectrum for 2014 (Fig.5), there is 
an unexpected comparative ``excess" of soft X-ray flux of the HD spectrum (once 
the high energy flux is renormalized to the fluxes of the NF spectrum) that is 
particularly evident.  This is to say, during the drop 
phase the soft flux exhibits less of a decrease than the hard flux in Fig.\,1.
The most likely realization of this puzzle is that nearly all 
the soft flux persists fully even when a substantial fraction of the
hard X-ray sources is occulted.
It follows that the soft and hard flux 
sources are not strictly cospatial. Rather, they are more
nearly outside the occulting lines of sight, and possibly distributed 
all around the star. Candidates for the origin of this flux are the warm
plasma components $kT_2$ and/or $kT_3$ mentioned in $\S$\ref{spmodl}.

The second unexpected result in both Figs.\, 4 and 5
is that at the ``ultra-soft" energies the slopes of the SD 
interval spectra became less steep, no longer diverging with spectral
fluxes from the {\em nominal flux} intervals.  In general, one expects
that the softest fluxes should be attenuated progressively toward softer
energies by photoelectric absorption. How can this result be explained? 
Reference to the model spectra of each of the component plasmas comprising 
the total spectrum of \gc,~ such Fig.\,7 of LSM10, Fig.\,5 of SLM12a, 
one sees that the warm
plasma emissions are no longer quite so  
subordinate to the hot plasma emissions as they are at higher 
energies. 
If, for example, the hot plasma responsible for the hard flux disappears
while residual warm plasma sites are still visible on the suface,
%over the limb decreases during the 
%eclipse of much of the hot plasma can peek through, particularly as 
%we just reported; 
%then as can be seen from the cited figures 
%if these plasmas lie elsewhere and
%are not occulted. In this situation 
the warm plasmas' emissions can have a greater influence at soft energies
%begin to ``peek through" the prevailing emission of the hot plasma 
and in particular cause a shallower spectral slope at the lowest energies.
We hasten to point out that this is a speculative conclusion based
on few independent spectral bins (albeit of two observations). More spectral
observations during extended SD events can test this explanation.

\subsection{The hard X-ray drops of \gc~ (and HD\,110432)}

Of special interest in Fig.\,1  
is the sustained drop in the hard X-ray curve
that occurs at 25-27\,ks into the observation and lasts for 5\,ks of the 
remaining 7\,ks of the observation. The soft/hard color curve is essentially
constant during this duration. Such a drop has not been reported for \gc.~ 
The event cannot be confused with the very brief ($\leq$ 1\,ks) ``cessations" 
first noted by \citet[][]{Smith et al.1998b}
and subsequent authors. Although this drop behavior is
unique in the \gc~ record, it did remind us of a similar drop of nearly 50\%
in the hard X-ray flux of the \gc-analog HD\,110432, which was first reported
by \citet{Smith et al.2012b}
and is reproduced as our Figure\,6, along with the soft/``all"
color curve. Here the ``soft" band is taken as 0.6-2\,keV and ``all" as
0.6-12\,keV.  The drop is similarly sudden and lasts for the remaining
40\,ks of the time series. As in Fig.\,1  
there is no accompanying change in the color curve although the drop
in integrated energies is halved. 
the time over which the hard flux is reduced with respect to the Fig.\,1
drop for $\gamma$\,Cas is halved or less.  

%FIGURE6
\begin{figure} 
%\label{hd2007}
\begin{center}
\vspace*{-0.00in}
\includegraphics*[width=6.2cm,angle=90]{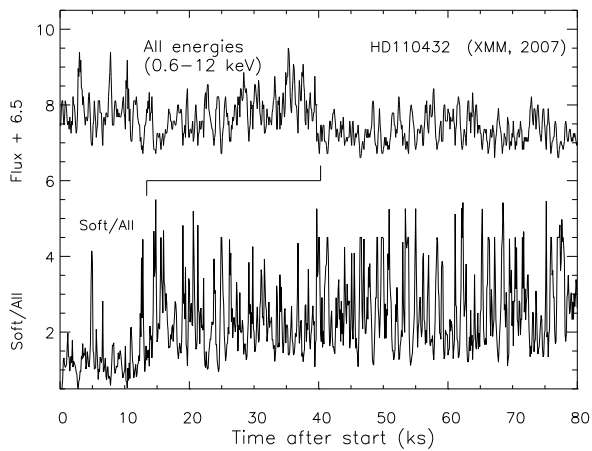}
\end{center}
\caption{
Upper and lower panels depict the All energies \xr\,\,and soft/hard color 
curves for HD\,110432 during the {\it XMM} observation of 2007 Sep 7; ``soft"
refers to 0.6-2\,keV fluxes.  A flat-bottomed soft dip occurs at 0-13\,ks. 
A hard dip, dropping by $\sim$50\%
from the nominal level before time 25\,ks, occurs some 27\,ks later. }
\end{figure}

  There is again to be an ambiguity in attributing these events to
changes in the source energy supply versus those in geometry.  On one 
hand, there are no other observational precedents for {\em sustained} drops
in the source term (impinging electron beams in the interaction scenario), 
although new surprises keep arising 
with increased observational scrutiny of these star's high energy behavior. 
On the other hand, we point out that the centroid of the soft dip for \gc~
in Fig.\,1 occurs 25\,ks before the transition of the hard drop. 
There is a corresponding interval of 27\,ks for the light and color curves 
of HD\,110432. In both cases this interval happens to be about $\frac{1}{4}$
of the Be star's rotational period (not precisely known). Thus, one 
explanation, consistent with the ``pencil column" explanation for the UV 
continuum dip in early 1996 (S19), is that a pre-wind absorption column 
is responsible for the soft dip. 
Then, the activity center disappears with rotation over
the receding limb of the Be star and approximately half the hard
X-ray flux is quickly removed from our visibility. There are doubtless
other explanations, and until further examples come to light we will not
speculate further on the ``hard drop phenomenon."

\subsection{Notes on the WD accretion hypothesis}
\label{mdtt}

  Limits on the secondary star's mass of the $\gamma$\,Cas system inferred 
from its mass function, as well as the optically thin, thermal spectrum, 
make it unlikely that a neutron star emits the X-rays.  
Also, limits on the contribution 
to UV flux rule out the secondary being a sdBO star. The secondary is either a
main sequence star, or more likely if $\gamma$\,Cas is a blue straggler, a white
dwarf.  A key difficulty for WD accretion, as noted long ago by \citet[][]{White 
et al.1982}, is for the Be star to have sufficient mass loss to power 
the observed L$_{x}$ of $\sim$10$^{33}$ erg\,s$^{-1}$ via WD accretion.

  Fitting of asymmetric wings of UV resonance line profiles led
\citet[][]{Hammerschlag-Hensberge et al.1980} to a mass loss estimate of
1$\times$10$^{-8}$ M$_{\odot}$\,yr$^{-1}$ for $\gamma$\,Cas. However,
a recent Viscuous Decretion Disk model, with fitting across the 
near infrared, has placed the loss rate of \gc~at 
2-2.5$\times$10$^{-10}$ M$_{\odot}$\,yr$^{-1}$ \citep[][]{Vieira et al.2017}.
This is up to three orders of magnitude below the range obtained by
\citet[][]{Lamer&Waters1987} of up to 5$\times$10$^{-7}$ 
M$_{\odot}$\,yr$^{-1}$), which is
near the value used by H16 to justify a WD accretion picture. 
We also point out that the Lamers \& Waters estimates were based on highly
uncertain assumptions, including an extrapolation of an arbitrary wind 
velocity at the star's surface as well as a large range in the assumed 
ionization equilibrium in the wind.  

Finally, contrary to H16, there is no evidence for clumping in the 
$\gamma$\,Cas wind on rapid timescales \citep[e.g.,][]{Cranmer et al.2000}, 
as could otherwise be reasonably argued for O and B supergiants
\citep{Lepine&Moffat2008,Prinja & Massa2010}. 
Altogether, these arguments suggest the WD accretion picture cannot
be sustained from what we now believe are the wind conditions for \gc.~

\section{Summary and Conclusions}

  The highlighting by \citet{Hamaguchi et al.2016} of the 
soft-energy X-ray dips (SDs), accompanied by minor or no hard X-ray 
dips, has revealed a new phenomenon by which to study the environment 
of the hard  X-ray  production of $\gamma$\,Cas. It is important to study 
this environment, first, to elucidate the locations and potentially the
creation mechanism of the hot plasma. Second, these studies reveal
a means to disambiguate the positions of the dominant ``hot" plasma
from the ``warm" ones. 

 Our estimates of particle densities, column densities, and sizes of the
soft X-ray absorbing ``canopies" are close to independent estimates of the
``cloudlets" responsible for {\it msf} in primarily UV lines.
In addition, the n$_{H}$ values for our ensemble of SDs exhibit a range
of 0.23-0.87$\times$10$^{22}$\,cm$^{-2}$, which perhaps fortuitously almost
perfectly matches the range 
0.24-0.81$\times$10$^{22}$\,cm$^{-2}$ found by H16 using {\it Suzaku} data. 
These values have a range of a factor of about four, suggesting 
that the canopies responsible for them have a distribution of sizes.
By comparison, these values are nearly two orders of magnitude higher than
the interstellar medium column estimate of 
1.45${\pm 0.3}$$\times$10$^{20}$\,cm$^{-2}$
from the most detailed UV data \citep[][]{Jenkins2009}.

Our analyses of spectra obtained from the SDs in 2014 and 2010-4 
indicates that the steep drop off in slope moderates at energies below 
0.7\,keV. This similarity suggests that although the emission of the hot 
plasma at 14\,keV dominates, it dominates less at these energies. 
The presence of warmer plasm components discussed by previous authors, 
which from our results appears to be spread more evenly over the star, 
is making its presence felt at these low energies.

We have also discovered, in agreement with the H16 and H18 studies, 
that the SDs can occur as more or less isolated light curve features 
events several hours apart (2014, 2010-2, and 2010-4) or as several 
brief events at seemingly random, but more closely spaced, intervals. 
Accumulating more statistics on the strengths and event-to-event 
frequencies of SD events in the future should help decide whether 
lines of sight to one surface center, per rotational advection, 
intersect to other nearby centers. In such a case a relatively large
number of centers - {\em or}, that there are few centers being ``turned
on and off" by repeated beam injections into those few centers.
Our preference at the moment is for repeated beam injections to produce 
most events, such as in the 2014 light curve. However, we suspect also 
the former case occurs sometimes, e.g., again as in the 2004 light curve. 
A resolution of this question may shed light on the degree and rate
of change of the magnetic topology, which in turn is critical to 
reconciling so far unsuccessful attempts to detect complex
magnetic fields by spectropolarimetric means.

  Finally, we have suggested that hard X-ray drops, unaccompanied by SDs,
can occur in both $\gamma$\,Cas and HD\,110432. More long observations
of these stars should be gathered to explore the idea that these
occur one-quarter of a rotational cycle after (and equally {\em before})
the appearance of a strong SD. Such a possibility would tie the existence
of major activity centers to the Be star's rotation. If this idea is
not born out, an alternate explanation must be advanced to understand
how much of both the soft and hard X-ray fluxes can be suddenly turned off 
for intervals of at least 2-3 hours. \\

 We dedicate this paper to the memory of Mrs. Eleanor Smith.


%%%%%%%%%%%%%%%%%%%%%%%%%%%%%%%%%%%%%%%%%%%%%%%%%%
%%%%%%%%%%%%%%%%%%%%%%%%%%%%%%%%%%%%%%%%%%%%%%%%%%
\bsp	% typesetting comment
\label{lastpage}
\end{document}